\documentclass[12pt]{article}%
\usepackage{amsmath}
\usepackage{graphicx}
\usepackage{amsfonts}
\usepackage{amssymb}%
\providecommand{\U}[1]{\protect\rule{.1in}{.1in}}
\begin{document}

\title{Does Lorentz Force Law Contradict the Principle and Theories of Relativity for
Uniform Linear Motion?}
\author{C. S. Unnikrishnan\thanks{E-mail address: unni@tifr.res.in}\\\textit{Gravitation Group, Tata Institute of Fundamental Research, }\\\textit{Homi Bhabha Road, Mumbai - 400 005, India}}
\date{}
\maketitle

\begin{abstract}
I show that no force or torque is generated in cases involving a charge and a
magnet with their relative velocity zero, in any inertial frame of reference.
A recent suspicion of an anomalous torque and conflict with relativity in this
case is rested. What is distilled as `Lorentz force' in standard
electrodynamics, with relative velocity as the parameter, is an
under-representation of two distinct physical phenomena, an effect due to
Lorentz contraction and another due to the Ampere current-current interaction,
rolled into one due to prejudice from special relativity applied only to
linear motion. When both are included in the analysis of the problem there is
no anomalous force or torque, ensuring the validity of Poincare's principle of
relativity. The issue of validity of electrodynamics without the concept of
absolute rest, however, is subtle and empirically open when general
noninertial motion is considered, as I will discuss in another paper.

\end{abstract}

The modern theory of electrodynamics was built by Maxwell by distilling
various experimental results and then including the phenomenon of displacement
current he discovered from a consistency analysis. This was well before the
theories of relativity by Lorentz, Poincar\'{e} and Einstein came into
discussion. The relativity principle expounded by Poincar\'{e} ensures the
impossibility of detecting uniform linear motion by experiments in
electrodynamics and mechanics \cite{Poincare}. Thus the state of uniform
rectilinear motion is equivalent to the state of rest. It was Einstein who
asserted, in the opening paragraph of his 1905 special relativity paper
\cite{Einstein1905}, that \emph{the reciprocal electrodynamic action of a
magnet and a conductor depends only on the relative motion of the conductor
and the magnet}. See figure 1 for the relevant situation. He further wrote,
"examples of this sort, together with the unsuccessful attempts to discover
any motion of the earth relatively to the \textquotedblleft light
medium\textquotedblright, suggest that the phenomena of electrodynamics as
well as of mechanics possess no properties corresponding to the idea of
absolute rest".%

\begin{figure}
[ptb]
\begin{center}
\includegraphics[
height=0.934in,
width=4.7331in
]%
{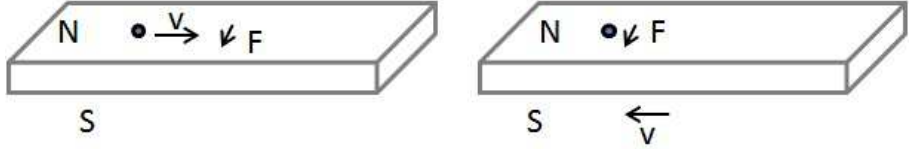}%
\caption{The charge moving at velocity $\vec{v}$ is physically \ equivalent to
magnet moving at velocity $-\vec{v}$ in Einstein's special relativity. Both
forces are given by the Lorentz force law with $\vec{v}$ as their relative
velocity.}%
\end{center}
\end{figure}

While analyzing the cases involving a conductor and magnet in various states
of relative motion between them, I had realized that current electrodynamics,
as represented in Maxwell's equations and the Lorentz force law, is actually
an under-representation of the fundamental physical phenomena involved
\cite{Unni-grex-poster}. In particular, the Lorentz force law and its
relativistic transformations \emph{involving only the relative velocity}
between charges and magnets mix up two distinct physical phenomena into a
single expression that does not adequately represent the physical effects for
general motion involving non-rectilinear motion. Something has been lost in
translation by Maxwell and Lorentz, from the body of results arrived at by
painstaking experiments over decades by Faraday, Ampere, Weber and others.
However, that shows up only in the case of non-rectilinear motion and the
subject of this paper is the analysis that shows that there is no torque or
force when a charge and a magnet that are relatively stationary are observed
from a uniformly moving frame.

The physical phenomena that are relevant are (a) Ampere's current-current
interaction and (b) the modification of charge densities involved in the
effective equivalent currents in a magnet due to Lorentz contraction in
motion. \emph{These are two different physical phenomena, with different
physical origins and mathematical expressions, which coincide exactly for the
case of force on a charge moving uniformly relative to a magnet.} Usually, the
force is written as the Lorentz force,
\begin{equation}
\vec{F}=q(\vec{E}+\vec{v}\times\vec{B})
\end{equation}
Since $\vec{v}$ is interpreted as the \emph{relative velocity} between the
magnet and the charge, there is no force when a charge is at rest relative to
a magnet and no force appears when the system is viewed from another frame
moving relative to the physical system. However, instead of using the formula,
if one tries to calculate the force from the interactions of the charges and
currents involved in the problem, there could be an apparent conflict that is
spurious if one of two fundamental physical effects we discussed is not taken
into account. Indeed, the current-current interaction is often forgotten in
discussions since it is assumed that Maxwell's equations contain fully its
manifestations. This is perhaps true, except that there is certainly one
ambiguous element in the formulation of modern electrodynamics, and that is
the interpretation of the velocity that appears in the transformation
equations: \emph{relative or absolute}? When the Lorentz force law is
discussed in moving frames, usually only the Lorentz contraction effect is
considered. In particular, the case of a charge and a magnet there is a
spurious force or torque unless the current-current interaction is included
\cite{Unni-grex-poster}. \ This seems to be the case in a news and analysis
article announcing a serious conflict between relativity and the Lorentz force
law, in the case of a charge and a magnet that are relatively stationary, when
observed from a moving frame \cite{science-news}. The magnet is equivalent to
a current loop involving two opposite currents separated by distance. In a
moving frame Lorentz contraction generates asymmetric charge densities and an
electric field (see below). \ If this is the only effect that is considered as
the basis of Lorentz force, an anomalous torque is indicated and if true, it
would conflict with relativity since a real force or torque and resultant
motion (dynamics) cannot appear within this physical system merely due to
uniform motion of an observer.

Now I show that this paradox arose precisely because of not taking into
account the fundamental phenomenon of current-current interaction in the
moving frame. Including that effect makes the torque vanish. Indeed, the
system considered in the news analysis is a variant of an example I have
analyzed earlier with the difference that he was considering a system where
the charge is separated from the magnet and a torque is generated, whereas I
have been dealing with examples where the charge is right near the surface of
the large magnet \cite{Unni-grex-poster}. (This is because we are attempting
to make measurements on the dynamics of charges near moving magnets in various
configurations and the possibility of good measurements reduces with
increasing distance between the charge and the magnet). In the particular case
of linear motion, the analysis can be done either with relative velocities or
with velocities with respect to a preferred absolute frame and the results are
identical, reiterating the fact that no experiment in linear motion can be
used as a demarcating test between different theories of relativity obeying
the relativity principle. Three such theories are the Lorentz-Poincar\'{e}
relativity theory \cite{Poincare,Lorentz}, Einstein's special relativity
\cite{Einstein1905} and the new paradigm of Cosmic Relativity
\cite{cosrel-isi,cosrel} that I have been advocating in which all relativistic
effects are due to the gravitational effects of cosmic matter and cosmic frame
is a preferred frame perfectly replacing the old ether.

Consider a charged particle in close vicinity of a bar magnet with large polar
area, close to one of its poles. Moving the charge relative to the magnet will
make it deflect transversely and moving the magnet also will produce the same
physical effect (figure 1). The force is given in both cases, as in modern
electrodynamics, by the Lorentz force law,
\begin{equation}
\vec{F}=q(\vec{E}+\vec{v}\times\vec{B})
\end{equation}
with $E=0$ in this case. This is the example that guided Einstein to (hastily)
conclude that \emph{all electrodynamic phenomena} are completely symmetric in
relative motion. Now consider the entire system of the magnet and the charge
at rest. In special relativity, $\vec{v}$ is the relative velocity between the
magnet and the charge, as originally asserted by Einstein. When the relative
velocity is zero, there is no force on the charge and it remains at rest.
However, viewed from a frame moving at velocity $\vec{V}$, the magnet becomes
charged, with opposite charges on its two sides parallel to the motion of the
observer due to Lorentz contraction (figure 2).%

\begin{figure}
[ptb]
\begin{center}
\includegraphics[
height=1.0118in,
width=4.4408in
]%
{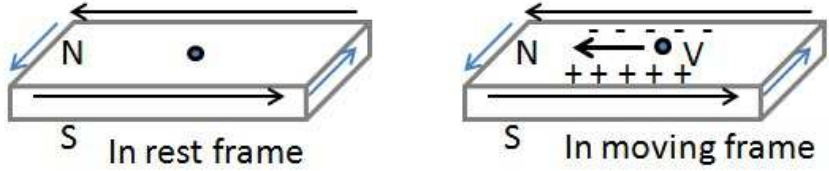}%
\caption{Lorentz contraction of currents in the magnet generates asymmetric
charge density and electric field. The long arrows depict the current
elements. In a frame moving right with velocity $\vec{V}$, the originally
stationary charge becomes a current with velocity $-\vec{V}$ .}%
\end{center}
\end{figure}
This is because the magnet is equivalent to a current loop and on one side the
velocity of the charged particles corresponding to the current add to the
motion of the observer and on the other side, it subtracts. If the speed of
the moving charges in the magnet is $u,$ the Lorentz contraction factors for
the two opposite currents are different, $\left(  1-\left(  V+u\right)
^{2}/c^{2}\right)  ^{1/2}$ and $\left(  1-\left(  V-u\right)  ^{2}%
/c^{2}\right)  ^{1/2}$. \ For low velocities involved for both the current and
the observer, these can be written as
\begin{align}
(1)\quad\Delta l_{1} &  =-L(V^{2}+u^{2}+2uV)/2c^{2}\nonumber\\
(1)\quad\Delta l_{2} &  =-L(V^{2}+u^{2}-2uV)/2c^{2}%
\end{align}
and the difference for low velocities is $-2LVu/c^{2}.$ (the similarity to the
Sagnac term is not accidental. See reference \cite{cosrel-isi,ijmpd-electro}).
The change in the linear density will be then $-2Vu/c^{2}.$ This multiplied by
the negative charge density $\lambda$ is the difference in charge densities,
creating a transverse electric field $\vec{E}$ in the direction $\vec{V}%
\times\left(  \mathbf{\nabla}\times\vec{u}\right)  ,$ of magnitude
\begin{equation}
E=2\lambda Vu/2\pi\varepsilon_{0}rc^{2}%
\end{equation}
However, now we have a conflict with principle of relativity if there is only
the Lorentz contraction, since the charge should start moving due to this
electric field, which is unphysical since an observer's mere motion should not
physically displace the charge relative to the magnet! \emph{However, note
that in the same frame, the charge becomes a current} $-q\vec{V}$ and there
are forces of opposite sign to that of the Lorentz contraction from the Ampere
current-current interaction that exactly bring the net force to zero. Parallel
current attract and opposite ones repel, where one current $i_{q}=qV$ is from
the moving test charge and the other two oppositely directed ones $i_{m}$ from
the current loop in the magnet. This current-current force is exactly same in
magnitude and opposite in direction to the one generated by the Lorentz
contraction. The two cancel, and the charge remains at rest, as if there is no
force. The Ampere interaction from the two opposite currents on the current of
the charge $q,$ now moving with velocity $\vec{V}$ in the new reference frame
is
\begin{equation}
F_{A}=\frac{2\mu_{0}}{2\pi}\frac{i_{q}i_{m}}{r}=\frac{\mu_{0}}{\pi}%
\frac{qVi_{m}}{r}=\frac{\mu_{0}}{\pi}\frac{qV\lambda u}{r}%
\end{equation}
and the (oppositely directed) force from Lorentz contraction and the electric
field of the modified charge density of the currents in the magnet is
\begin{equation}
F=qE=\frac{q\lambda Vu}{\pi\varepsilon_{0}c^{2}r}=\frac{\mu_{0}}{\pi}%
\frac{qV\lambda u}{r}=\frac{\mu_{0}}{\pi}\frac{i_{q}i_{m}}{r}%
\end{equation}
if we identify $\mu_{0}=1/\varepsilon_{0}c^{2},$ which is the case in
electrodynamics! If the charge, instead of being near the pole of the magnet,
where the field is uniform, is far away as illustrated in reference
\cite{science-news} then we get a (spurious) nonzero torque from just the
Lorentz contraction (figure 3).%

\begin{figure}
[ptb]
\begin{center}
\includegraphics[
height=1.011in,
width=5.1024in
]%
{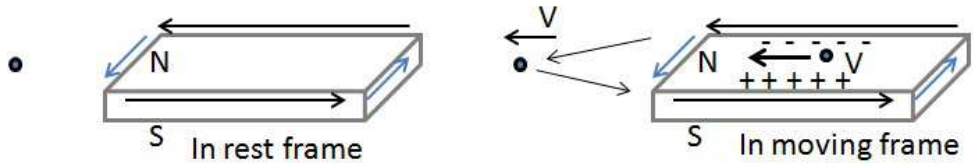}%
\caption{Spurious torque on the magnet due to a charge on the far left,
considering Lorentz contraction alone in the problem in the moving frame. The
current-current interaction cancels this.}%
\end{center}
\end{figure}
The problem posed as a paradox conflicting relativity is completely solved if
the torque is recalculated with the current-current interaction added. That
generates an equal and opposite torque to what was calculated from Lorentz
contraction and the net torque is zero. The factual situation is that there
are two equal and opposite forces (or forces and torques in the general case),
which balance in the case of linear motion. This should not be trivialized, as
is done in standard electrodynamics, to a `zero force' situation since this
balance is broken in the case of rotation motion as we shall see
\cite{unni-unipolar}. A `zero' as genuine nothing and as a delicate and
special balance of two equals are completely different, conceptually and physically.

It is clear that, for the case depicted in fig. 3, one may take the limit of
the current loop area shrinking to infinitesimal values with the magnetic
moment remaining constant and the analysis, with the current-current
interaction included, gives zero torque in all frames in the presence of the charge.

I may add here is that what we have done is to derive the relation between
$\mu_{0},\varepsilon_{0}$ and $c$ from the principle of relativity with
Ampere's current-current interaction and the Lorentz contraction as the
ingredients, without entering the plane of the electrodynamic wave equation.
It is very important to recognize that these phenomena of linear motion does
not allow us to demarcate between special relativity and an absolute frame
theory as the empirically valid theory because both incorporate Poincar\'{e}'s
principle of relativity. Explanations can be found for all electrodynamic
phenomena in linear motion in both theories by interpreting the velocities
involved as absolute velocities in absolute frame theories, and as relative
velocity between the magnet and the conductor in special relativity. Both
predict zero force and zero torque in the problems discussed above.

Even in the simpler problem of a single conductor and an external charge, the
paradox appears if the current-current interaction is ignored. But such
anomalies are spurious since in the same moving frame the charge is
transformed to a current that interacts with the current in the conductor
cancelling the force from Lorentz contraction. Principle of relativity in
rectilinear motion remains intact. However, the respite is temporary, as I
will elaborate in another paper dealing with cases of uniform rotational
motion~\cite{unni-unipolar}. It is also important to recognize that the wise
do not consider that modern electrodynamics as discussed in standard text is
the last word on the electrodynamics of the real world and several issues
related to the exact nature of the current-current interaction, unipolar
induction, action and reaction etc. are being actively discussed and
experimented with, albeit in a relatively small yet competent community.

\bigskip

\noindent{Disclaimer: The discussion in this paper pertains to the news and
analysis, reference [4], and not directly to the original paper, which was not
available at the time of writing or submitting this paper. The paper by M.
Mansuripur is now\ available in arXive.org as }arXiv:1205.0096.


\begin{thebibliography}{9}                                                                                                %


\bibitem {Poincare}H. Poincar\'{e}, \textit{L'\'{e}tat actuel et l'avenir de
la physique math\'{e}matique}, Bulletin des sciences math\'{e}matiques 28 (2):
302-324 (1904), English translation available online:

http://archive.org/details/foundationsscie01poingoog

\bibitem {Einstein1905}A. Einstein, \textit{Zur Elektrodynamik bewegter
K\"{o}rper,} Annalen der Physik \textbf{322} (10): 891 (1905), \textit{On the
Electrodynamics of Moving Bodies}, English translation by M. Saha in The
Principle of Relativity: Original Papers by A. Einstein and H. Minkowski,
University of Calcutta, 1920 (available online).

\bibitem {Unni-grex-poster}C. S. Unnikrishnan, \textit{New experiments and
results on the relative velocity of light and on the electrodynamics of
currents, and their links to gravity, \ }Poster at GPhys Kick-off symposium,
\ Les Houches 2009, \ http://gphys.obspm.fr/LesHouches2009/Program.html

\bibitem {science-news}A. Cho, \textit{Textbook Electrodynamics May Contradict
Relativity, }Science \textbf{336}, 404 (2012) based on M. Mansuripur,
\textit{Trouble with the Lorentz law of force: Incompatibility with special
relativity and momentum conservation, }to appear in Phys. Rev. Lett.
(2012)\textit{; }arXiv:1205.0096.

\bibitem {Lorentz}H. A. Lorentz, \textit{Electromagnetic Phenomena in a System
Moving with Any Velocity Less Than That of Light}, Proc. Acad. Science
Amsterdam \textbf{IV}, 669 (1904).

\bibitem {cosrel-isi}C. S. Unnikrishnan, \textit{Physics in the `Once-Given'
Universe,} in Recent Developments in Theoretical Physics, p 99, (Eds. Subir
Ghosh and Guruprasad Kar, World Scientific, 2010).

\bibitem {cosrel}C. S. Unnikrishnan, \textit{Cosmic Relativity: The
Fundamental Theory of Relativity, its Implications, and Experimental Tests},
gr-qc/0406023 (this version contains some errors and a misjudgment,
specifically in the section discussing the velocity of light. However, the
gist of the theory is accurate. \ One-way relative velocity of light cannot be
a constant in an absolute frame theory).

\bibitem {ijmpd-electro}C. S. Unnikrishnan and G. T. Gillies,
\textit{Gravito-Electromagnetism}: \textit{Glimpses\ Glimpses of Unexplored
Connections}, \ Int. Jl. Mod. Phys. D \textbf{13}, 2321 (2004).

\bibitem {unni-unipolar}C. S. Unnikrishnan, \textit{Complete Solution of the
Fundamental Problems of Unipolar Induction and Related Phenomena, }in
preparation\textit{.}
\end{thebibliography}
\end{document}